\documentclass[prl,floatfix,twocolumn,showpacs,amsmath,amssymb]{revtex4}
\usepackage{dcolumn}
\usepackage{bm}
\usepackage{graphicx}
\usepackage{amsfonts}
\usepackage{amsmath}
\usepackage{amssymb}

\begin{document}

\title{From the triangular to the kagome lattice: Following
the footprints of the ordered state.}

\author{Liliana Arrachea,$^{1,2,*}$ Luca Capriotti,$^3$ and Sandro Sorella$^1$}
\affiliation{
$^{1}$ International School for Advanced Studies (SISSA) and INFM-Democritos, National Simulation Centre, I-34014 Trieste, Italy \\ 
$^{2}$ Max Planck Institut f\"ur Physik komplexer Systeme, Dresden, Germany \\ 
$^{3}$  Kavli Institute for Theoretical Physics, University of California, Santa
 Barbara CA 93106-4030, USA 
}

\date{\today}

\begin{abstract}
We study the spin-$1/2$ Heisenberg model in a lattice
that interpolates between the triangular and the kagome lattices.
The exchange interaction along the bonds of the
kagome lattice is $J$, and the one along the bonds connecting kagome and
non-kagome sites is $J'$, so that
$J^\prime=J$ corresponds to the triangular limit and $J^\prime=0$ to the 
kagome one.  
We use variational and exact diagonalization techniques.
We analyze the behavior of the order parameter for the
antiferromagnetic phase of the triangular lattice, the spin gap, and the
structure of the spin excitations as functions of $J^\prime/J$. 
Our results indicate that the antiferromagnetic order is not affected
by the reduction of $J'$ down to $J^\prime/J\simeq 0.2$. 
Below this value, antiferromagnetic correlations grow weaker, a description  
of the ground state in terms of a N\`eel phase renormalized by
quantum fluctuations becomes inadequate, and the finite-size spectra
develop features that are not compatible with antiferromagnetic ordering.
However, this phase does not appear to be connected 
to the kagome phase as well, as
the low-energy spectra do not evolve with continuity for $J^\prime\to 0$
to the kagome limit. In particular, for any non-zero value of 
$J^\prime$, the latter interaction sets the energy scale for the 
low-lying spin excitations, and a gapless triplet spectrum,
destabilizing the kagome phase, is expected.
\end{abstract}

\maketitle

\section{Introduction}
Geometrically frustrated antiferromagnets are the historical
candidates for the realization of a {\em spin liquid} ground state.
Indeed, the spin-half
Heisenberg antiferromagnet on the triangular lattice
was the first model to be proposed by Anderson and Fazekas \cite{an,fa-an} 
in 1973 as a system where geometric frustration
and quantum fluctuations could prevent 
zero-temperature magnetic ordering in two dimensions, 
stabilizing instead a ground state with gapped spin excitations 
and exponentially decaying correlations.
Since then, a good amount of work has been devoted to investigate
the nature of the ground state of the triangular Heisenberg model
\cite{huse,tri1,bernu1, bernu2,tri2,tri3,tri4,tri5,aza,luc,luct,boni} which
remained an open question until quite recently. At present, however,
there is a general consensus on the existence of long-range
antiferromagnetic order following a $120^\circ$ N\`eel pattern in the
ground state of this model\cite{bernu1,bernu2,luc,boni}:  frustration
and quantum fluctuations on the two-dimensional triangular lattice are not
strong enough to stabilize a non-magnetic ground state.

\begin{figure}[t]
\includegraphics[width=80mm]{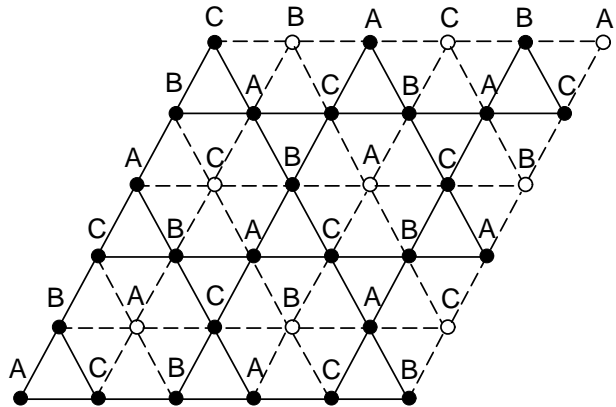}
\caption{The depleted triangular lattice. Filled and empty circles
are the kagome and non-kagome sites; solid and dashes
lines indicate $J$ and $J^\prime$ bonds of the Hamiltonian 
(\ref{ham}). The letters A,B,C label the three different spin directions 
oriented $120^\circ$ apart of the $\sqrt{3}\times\sqrt{3}$ classical
N\`eel state.} \label{fig1}
\end{figure}

A more promising candidate for a disordered ground state can be 
obtained through a `dilution' of the triangular lattice, 
leading to the so-called {\em kagome} net (Fig.~\ref{fig1}). 
In fact, on this geometry, 
due to the lower coordination ($z=4$ compared with $z=6$ in the 
triangular case), frustration is much stronger and even
in the classical limit it gives rise to an infinite 
number of classical ground states, with ordered and disordered configurations
degenerate in energy.\cite{kagoc1,kagoc2,kagoc3}
Due to the extensive entropy of the classical ground state, 
the so-called {\em order from disorder} mechanism -- usually stabilizing, 
among degenerate manifolds,  long-range ordered configurations --
is much less effective than in other frustrated models. 
In particular, while harmonic fluctuations select planar configurations, they
turn out to be completely insensitive to their degree of order,
and only non-linear effects eventually stabilize a 
classical ground state with  N\`eel correlations
with a $\sqrt{3}\times\sqrt{3}$ pattern (see Fig.~1). 
Whether such classical minimum energy configuration possesses
true long-range order \cite{kagoc1} or it is a critical point with power-law 
correlations \cite{kagoc2,kagoc3} is still an open issue. In any case, 
due to the particularly delicate mechanism leading to the
classical antiferromagnetic order, the latter is expected to be easily
destabilized by quantum fluctuations.

The investigation of the quantum Heisenberg antiferromagnet
on the kagome lattice has been capturing increasing attention 
\cite{kago1,kago2,kago3,kago4,kago5,kago6,kago7,lech,wal,mam,t2,lhul} 
for some time now. Mainly on the basis of numerical
work, evidence has accumulated supporting a spin-liquid ground state,
even though a very peculiar one: this, in fact, would be 
characterized by a small gap ($\sim J/20$) to spin excitations, and by 
an exponentially large number of singlets contiguous to the ground  
state.\cite{lech,wal,mam,t2,lhul}
The classification of spin liquid of
Type II has been recently proposed \cite{t2} to  classify
this particular behavior.

 On the experimental side, the triangular geometry provides the
scenario for interesting physical phenomena taking place in 
 the superconducting compounds
Na$_x$CoO$_{2-y}$H$_2$O \cite{trisu} and the organic materials
$\kappa$-(BEDT-TTF)$_2$X, with X being I$_3$, Cu[N(CN)$_2$]Br
or Cu(SCN)$_2$ \cite{orga}.
 Two spin-1/2 kagome-like materials have been also recently reported, 
the Volborthite Cu$_3$V$_2$O$_7$(OH)$_2$ 2H$_2$O \cite{vol} and
the kagome-staircase compounds Ni$_3$V$_2$O$_8$ and Co$_3$V$_2$O$_8$
\cite{kast}.

In the present work, we consider the spin-$1/2$ Heisenberg antiferromagnet
on a lattice that interpolates between the triangular and the
kagome ones. The Hamiltonian is
\begin{equation}
{\hat H}=J \sum_{\langle ij\rangle} \hat{{\bf S}}_i \cdot {\hat{\bf S}}_j  +
J^{\prime} \sum_{\langle i j'\rangle} \hat{{\bf S}}_i \cdot {\hat {\bf S}}_{j'},
\label{ham}
\end{equation}
where $\hat{{\bf S}}_i$ are spin-half operators,
$\langle ij\rangle$ denotes nearest-neighbor bonds belonging to the
kagome lattice, and $\langle ij'\rangle$ are the remaining
bonds connecting kagome and non-kagome sites. 
A scheme is indicated in Fig.~\ref{fig1}.
In this way, $J'=J$ corresponds to the usual triangular lattice,
while $J'=0$ defines the kagome one. Our aim is to start from
the triangular limit and to investigate the stability of the
ordered state as $J'/J$ decreases. This is sensible because 
the classical N\`eel state on the triangular lattice is 
compatible with the expected $\sqrt{3}\times\sqrt{3}$ 
classical ordering on the kagome antiferromagnet.
The first to study this model were Zeng and Elser \cite{kago5},
who performed a spin-wave analysis and concluded that, 
for spin-1/2 particles, the ordered state could be stable 
for $J'$ down to $J'/J\simeq 0.2$. Very recently, this model has been
investigated with a coupled cluster treatment \cite{farn}. This
technique is based on the three-sublattice structure
characterizing the $120^o$-N\'eel order of the triangular lattice,
which is found to break down very close to $J'=0$, indicating the
instability of the magnetic ordered state very close to the 
kagome limit.

In this paper, we tackle this problem using variational approaches,
and exact diagonalization of small clusters. In particular, in
Section II, we employ the so-called fixed-node (FN)  technique \cite{fn}
in order to improve the accuracy of a wave function with long-range
antiferromagnetic order previously introduced in the pure
triangular case \cite{huse,luc}. This technique, is usually used
in the context of quantum Monte Carlo simulations as a method 
to approximate the  Hamiltonians affected by sign-problem 
instabilities \cite{fn} and obtain exact ground-state properties 
of the corresponding ``effective Hamiltonian'', no longer affected by the 
sign problem. Here we use the FN  method 
to define a variational state with long-range antiferromagnetic 
order, and we check its accuracy in describing the 
ground state of the model as $J^\prime/J$ is reduced
through a direct comparison with
exact diagonalization results on the $6\times6$ cluster.
In order to detect any indication of a change in the nature of the ground
state approaching the kagome limit, in Section III we analyze the structure of 
the low-energy spectra as a function of $J^\prime/J$, using the exact 
diagonalization of the Hamiltonian on small clusters. 
Section IV is finally devoted to summary and conclusions. 

\section{Variational approaches} 

A fairly accurate representation of the ground state of the spin-$1/2$
Heisenberg antiferromagnet on the triangular lattice can be obtained starting
from a $120^\circ$ N\'eel ordered state and including Gaussian fluctuations
by means of a Jastrow factor containing
two-spin correlations\cite{huse,luc}
\begin{equation}
|\psi_v\rangle= \hat{P}_0 \exp \Big( \frac{1}{2} \sum_{i,j}
v(i-j) \hat{S}^{z}_{i} \hat{S}^{z}_{j} \Big) |N\rangle~,
\label{wf}
\end{equation}
where $\hat{P}_0$ is the projector onto the $S^{z} = 0$ subspace, 
$|N\rangle$ is the classical N\'eel state in the $xy$ plane,
\begin{equation}
|N\rangle = \sum_{x} \exp{\Big[\frac{2\pi i}{3} 
\Big( \sum_{i \in {\rm B}} S_{i}^z - \sum_{i \in {\rm C}} S_{i}^z \Big) \Big]} |x\rangle~,
\end{equation}
and $|x\rangle$ is an Ising spin configuration specified 
by assigning the value of $S^z_i$ for each site. 
On the square lattice case, the classical N\'eel state reproduces exactly the 
phases of the ground state of the Heisenberg Hamiltonian according to the 
Marshall theorem.\cite{luc} On the triangular lattice, instead, the exact
phases of the ground state are unknown, and the classical part of the
wave function (\ref{wf}) does not reproduces them accurately. However,
as originally suggested by Huse and Elser, \cite{huse} a very accurate
ansatz of the ground-state phases can be obtained by including 
three-spin correlation factors of the form:
\begin{equation}
T(x) = \exp{\Big({\it i}\,\beta\sum_{\langle i,j,k\rangle}
\gamma_{ijk} {S}_{i}^z {S}_{j}^z {S}_{k}^z \Big)}~,
\label{eq.triplet}
\end{equation}
defined by the coefficients $\gamma_{ijk}=0,\pm 1$, appropriately
chosen so as to preserve the symmetries of the classical N\'eel state, and  by  the variational parameter
$\beta$. In particular the sum in Eq.~(\ref{eq.triplet}) runs over all
distinct triplets of sites $i,j,k$ where both $i$ and $k$ are nearest neighbors of
$j$, and $i$ and $k$ are next-nearest neighbors to one another.
The sign factor $\gamma_{ijk}=\gamma_{kji}=\pm 1$ is invariant under rigid translations and rotations
in real space by an angle of $120^\circ$ of the three-spin cluster 
$i,j,k$, but changes sign under rotations by $60^\circ$.
The resulting wave function reads therefore:
\begin{equation}
|\psi_v\rangle= \hat{P}_0 \sum_x \Omega(x)
\exp \Big(  \frac{1}{2} \sum_{i,j}
v(i-j) S^{z}_{i} S^{z}_{j} \Big) |x\rangle~,
\label{eq.wfhuse}
\end{equation}
with the phase factor given by
\begin{equation}
\Omega(x)=T(x) \exp{\Big[\frac{2\pi i}{3}
\Big( \sum_{i \in {\rm B}} S_{i}^z
- \sum_{i \in {\rm C}} S_{i}^z \Big) \Big]}~.
\end{equation}
Since the Hamiltonian is real, a better variational wave function
is defined by the real part of Eq.\ref{eq.wfhuse}.

The two-body Jastrow potential in (\ref{eq.wfhuse}) contains
in principle as many variational parameters as the independent
distances on the lattice. However, the same level of accuracy can be
obtained by optimizing separately the nearest-neighbor and next 
nearest-neighbor distances and adopting for the longer-range correlations
an expression based on the consistency  with linear spin-wave theory:
\cite{luc}
\begin{equation}
v(r)=\frac{\eta_{\infty}}{N}\sum_{q\ne 0} e^{-i q\cdot r} v_q
\label{jast}
\end{equation}
with 
\begin{equation}
v_q=1-\sqrt{\frac{1+2\gamma_k}{1-\gamma_k}}~,
\end{equation}
$\gamma_k=2(\cos{k_x}+2\cos{k_x/2}\cos{\sqrt{3}k_y/2})$ and 
$\eta_{\infty}$ is a variational parameter. 
For the anisotropic
triangular lattice (see Fig.\ref{fig1}) we have optimized separately the 
nearest-neighbor bonds connecting two kagome
sites 
($\eta_1$), 
and a kagome site and a non-kagome site
($\eta_1^\prime$), 
as well as the next-nearest-neighbor bonds
($\eta_2$). 
As a result, the total number of variational parameters for the
present variational wave function is 5. 
Their optimal values, and the corresponding variational energies, 
are reported for $N=36$ 
in Tab.~\ref{vartab}, for various values of $J^\prime/J$.

\begin{table}
\begin{tabular}{ccccccccc}
\tableline
\tableline
$J^\prime/J$ & $\beta$ & $\eta_1$ & $\eta_1^\prime$ & $\eta_2$ & $\eta_\infty$ &
$E_v/J$ & $E_0/J$ \\
\tableline
0.1        & 0.20    & -0.63     & -0.58            & 0.055     & 1.00    &
-10.7687 & -12.0799\\
0.2        & 0.20    & -0.63     & -0.58            & 0.055     & 1.00    &
-11.7299 & -12.7120  \\
0.3        & 0.20    & -0.63     & -0.58            & 0.055     & 1.00    &
-12.6911 & -13.5026\\
0.4        & 0.20    & -0.63     & -0.58            & 0.055     & 1.00    &
-13.6522 & -14.3708 \\
0.6        & 0.20    & -0.65     & -0.62            & 0.055     & 1.00    &
-15.5730 & -16.2287\\
0.8        & 0.23    & -0.70     & -0.69            & 0.055     & 1.00    &
-17.4873 & -18.1754\\
1.0        & 0.23    & -0.73     & -0.73            & 0.055     & 1.00    &
-19.4239 & -20.1734\\
\tableline
\end{tabular}
\caption{Variational parameters and variational energies
for the spin-wave wave function (\ref{eq.wfhuse})
for different values of the ratio $J^\prime/J$ on the $N=36$ cluster.
The Lanczos 
exact values of the energy are also reported for comparison. }
\label{vartab}
\end{table}

\begin{figure}
\includegraphics[width=80mm]{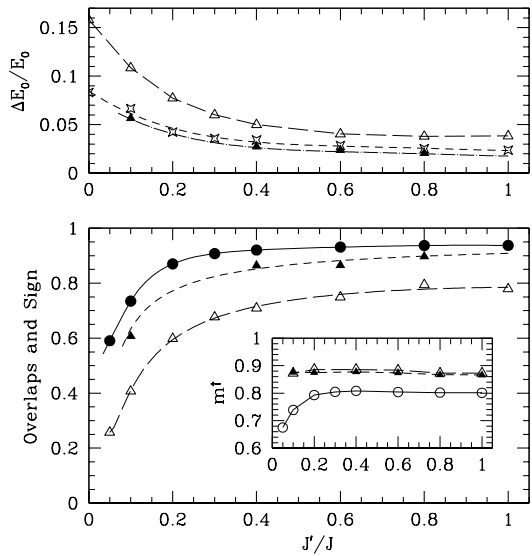}
\caption{Results on the $N=36$ cluster. Upper panel:
Relative error on the ground-state energy,
for the spin-wave (empty triangles)
and FN  (full triangles) wave functions. Stars refer to the
accuracy of the
upper bound on the energy given by the lowest
eigenvalue of the FN  Hamiltonian, $E_0^{\rm fn}$.
Lower panel: average sign of both the spin-wave and FN  wave function
(circles) and their overlap with the exact ground state (same symbols
as above). Inset: antiferromagnetic order parameter
The circles are the exact ground-state values, while empty and full
triangles correspond to the results obtained with spin wave 
and FN wave functions, respectively.
 Lines are guides for the eye.}
\label{fig2}
\end{figure}

The accuracy of the present long-range ordered
wave function in the triangular lattice limit
($J'/J=1$) has been analyzed in detail in
Ref.\cite{luc,luct}. In this limit the wave function is
known to provide a qualitatively correct representation of
the ground-state correlations.\cite{huse,luc,boni,luct}
In order to check the accuracy within a larger range of  $J^\prime/J$ 
we have compared 
several variational  properties with the exact ground-state values calculated
by exact diagonalization on the largest cluster presently accessible, $N=36$.
As shown in Fig.~\ref{fig2}, for $0.4 \lesssim J^\prime/J \le 1$ both the
accuracy on the ground-state energy and the overlap with the exact 
ground state,  remain approximatively constant and equal to the values in 
the triangular limit ($J^\prime/J=1$). 
In addition, the wave function (\ref{wf}) provides
an accurate representation of the phases of the actual 
ground state up to $J^\prime/J\simeq 0.2$, as it can be checked
by measuring the average-sign 
\begin{equation}
\langle s\rangle=\sum_x {\rm sgn}[\psi(x) \psi_0(x)] |\psi_0(x)|^2,
\label{sig}
\end{equation}
with $|\psi_0\rangle=\sum_x \psi(x)|x\rangle$ (Fig.~\ref{fig2}).
This remarkable feature is due to the
presence of the triplet term (\ref{eq.triplet}), allowing to 
adjust the phases in a non-trivial way, without changing
the underlying N\'eel order. For instance, in the triangular limit the
average sign (overlap) of the wave function is 0.733 (0.562) 
without the triplet term and 0.932 (0.779) with it. 

Since the variational wave function (\ref{wf}) reproduces accurately
the phases of the ground state its quality can be improved by adopting 
the FN  scheme of Ref.\cite{fn}. This allows one to obtain a new
variational wave function, $|\psi_{\rm fn}\rangle$, defined as the ground 
state of the so-called FN  effective Hamiltonian, whose matrix elements,
$H^{\rm eff}_{x,x^\prime}$, can be constructed starting from the original 
Hamiltonian and a variational guess on the ground-state phases given, 
in our case, 
by the wave function (\ref{eq.wfhuse}), $|\psi_v\rangle=\sum_x \psi_v(x)|x\rangle$: 
\begin{eqnarray} \label{fn}
H^{\rm eff}_{x^\prime,x}=\left\{
\begin{array}{ccl}
H_{x^\prime,x} & {\rm if} & \bar H_{x^\prime,x}\le 0 \\
0 & {\rm if} & \bar  H_{x^\prime,x}>0 \\
H_{x,x} + {\cal V}(x) & x=x^\prime
\end{array} \right.
\label{ofdi}
\end{eqnarray}   
where $\bar H_{x^\prime,x}=\psi_v(x^\prime)H_{x^\prime,x}/
\psi_v(x)$,
and 
\begin{equation} \label{eq.signflip}
{\cal  V}(x) =
\sum\limits_{\{\bar H_{x^\prime, x} > 0,\,\, x^\prime \ne x\}}
\bar{H}_{x^\prime,x}~.
\end{equation}

Indicating with 
$E^{\rm fn}_0$ the ground-state energy of the FN  Hamiltonian,
with $E^{\rm fn}_v = \langle \psi_{\rm fn} | \hat{H} |  \psi_{\rm fn}\rangle
/\langle \psi_{\rm fn} |\psi_{\rm fn}\rangle$ and 
$E_v = \langle \psi_v|\hat{H}|\psi_v\rangle/\langle
\psi_v|\psi_v\rangle$ the energy expectation values
on $|\psi_{\rm fn}\rangle$ and $|\psi_v\rangle$, respectively,
it is possible to show \cite{fn} that the following chain of inequalities holds:
$$E_v \ge E^{\rm fn}_0 \ge E^{\rm fn}_v \ge E_0~,$$
where $E_0$ is the ground-state energy of $\hat{H}$.
Hence, the FN procedure is granted
to produce a wave function with a better variational energy than
$|\psi_v\rangle$. 
In addition, also the lowest eigenvalue of the FN  Hamiltonian, 
$E^{\rm fn}_0$, gives an upper bound of the ground-state energy better 
than the variational energy $E_v$. 
This is the quantity which is usually considered
in the quantum Monte Carlo application since it is the most directly 
accessible.


The FN  Hamiltonian is explicitly defined in such a way that the
matrix $\bar H^{\rm eff}_{x^\prime,x}
= {\rm sgn} [\psi_v(x^\prime)] 
H^{\rm eff}_{x^\prime,x} {\rm sgn}[\psi_v(x)]$ has all negative 
off-diagonal matrix elements. By the Perron-Frobenius theorem, 
the amplitudes of the ground state of this matrix,
${\bar \psi}_{\rm fn}(x)$, have the same sign for all the configurations
$|x\rangle$.
This implies in turn that the FN  ground state, $|\psi_{\rm fn}\rangle$,
and the starting variational wave function, $|\psi_v\rangle$, have
exactly the same phases. In fact, the amplitudes of the
ground state of $H^{\rm eff}_{x^\prime,x}$, $\psi_{\rm fn}(x)$, are related
to the amplitudes of the ground state of the transformed matrix
${\bar H}^{\rm eff}_{x^\prime,x}$, ${\bar \psi}_{\rm fn}(x)$,
by the simple relation ${\bar \psi}_{\rm fn}(x)=\psi_{\rm fn}(x) 
{\rm sgn}[\psi_v(x)]$.
For this reason, the FN  wave function 
is expected to provide an accurate description of the ground state only  when
it is constructed with a good variational ansatz
of the ground-state phases. In the present case
for $0.2\lesssim J^\prime/J\le 1$.

In order to thoroughly check the accuracy of the FN  wave function,
we have exactly diagonalized with the Lanczos algorithm
the FN  Hamiltonian for several values of 
$J^\prime/J$ on the $N=36$ cluster. This has allowed us to
calculate not only the FN  energies but also the overlap
of the FN  wave function with the exact ground state.
As shown in Fig.~\ref{fig2}, both the FN  upper bounds to 
the ground-state energy, $E_v^{\rm fn}$ and $E_0^{\rm fn}$, 
are sizably more accurate 
than the simple variational estimate $E_v$. In particular, the FN  
wave functions has a much higher overlap than $|\psi_v\rangle$, and
its accuracy is almost constant and 
comparable to the one in the triangular limit down 
to values of $J^\prime/J$ as small as $\sim 0.2$. 

We have finally compared the variational, FN, and exact estimates of
the antiferromagnetic order parameter for a $120^\circ$ N\`eel order,
\begin{equation}
m^{\dagger 2}= 36 \frac{{\cal M}^2}{N(N+6)}~, 
\end{equation}
where ${\cal M}^2$ is the sublattice magnetization squared \cite{bernu2}.
Interestingly, by decreasing $J'/J$ the exact ground-state order parameter
remains approximatively constant down to $J'/J\simeq 0.2$ thus
indicating a possible destabilization of the antiferromagnetic order
only very close to the kagome limit. 
In addition, though the
variational and FN  estimates of the order parameter are approximatively
$10\%$ higher than the exact one the same degree of agreement is observed in
all the range $0.2\lesssim J^\prime/J\le 1$.
For the $N=36$ cluster, the FN  wave function, 
based on the  variational wave function (\ref{eq.wfhuse}), 
provides a good quantitative description of the exact ground state 
for $0.2 \lesssim J'/J \leq 1$. In this range, 
the variational and the exact expectation values of antiferromagnetic order 
parameter remain constant and equal to their values in the triangular limit.
Below $J'/J\simeq 0.2$, the exact value of the order parameter begins to 
decrease and  the accuracy of the our N\`eel ordered wave function 
quickly degrades, indicating a change in the ground-state correlations 
only very close to the kagome limit.

\begin{figure}
\vspace{-5mm}
\includegraphics[width=80mm]{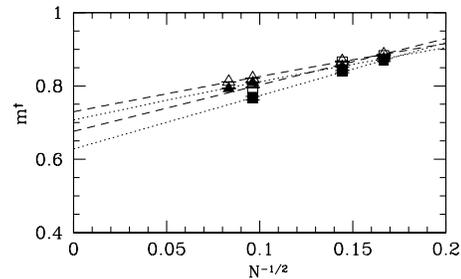}
\vspace{0mm}
\caption{Size scaling of the antiferromagnetic order parameter 
for the spin-wave (empty symbols and dashed lines), and the FN 
(full symbols and dotted line) wave functions: $J'/J=0.2$ (triangles),
$J'/J=1.0$ (squares).}
\label{opar}
\end{figure}

In order to support further  the stability of the antiferromagnetic phase 
for  very small values of $J^\prime/J$ we have extended the variational 
and the FN 
 calculations  of the order parameter $m^{\dagger}$ to much larger 
sizes by using quantum Monte Carlo techniques.\cite{calandra} 
As shown in Fig.~\ref{opar}, even 
at a value of $J^{\prime}/J=0.2$, the lowest coupling ratio when the 
variational wave function is expected to be accurate (see Fig.\ref{fig2}), 
the order parameter 
 $m^{\dagger}$  remains sizably larger than 
the triangular  $J^{\prime}/J=1$ case. This indicates that, as we increase the 
size at fixed ratio $J^{\prime}/J<1$, the stability of the  ordered phase, 
already evident in the exact diagonalization in the $N=36$ cluster 
(see Fig.\ref{fig2}), becomes more and more  clear. We expect that 
this qualitative behavior, confirmed both by the  exact diagonalization, 
and even more strongly  by  the variational and the FN  
approaches  on larger sizes, is a genuine feature of the model, 
even though the quantitative results that we have obtained 
 by quantum Monte Carlo may be affected by a sizable error. 
This feature  may appear  rather surprising, 
  as the quantum fluctuations should 
increase for small $J^{\prime}/J$ and should tend to destabilize the ordered 
phase, as expected for instance within spin-wave theory.\cite{kago1} 
 However, the wave function that we have used is consistent with spin wave 
theory in the large spin  limit, and therefore, since also at the variational 
level the value of $m^{\dagger}$ increases, we conclude that the quantum 
fluctuations are not very  accurately described within a method that is 
not controlled by the variational principle (the large spin limit), at least 
in the region of small $J^{\prime}/J$.   

\begin{table}
\begin{tabular}{lcccccc}
\tableline
$J^\prime/J$ & 0.0 & 0.2    &  0.4    &  0.6    & 0.8    & 1.0     \\
\tableline
$e_v$     & -0.369  & -0.324 &  -0.374 &  -0.425 & -0.478 &  -0.532 \\
$e^{\rm fn}_0 $ & -0.392  & -0.334 &  -0.381 &  -0.431 & -0.482 &  -0.537 \\
Ref.~\cite{farn} ($m=6$) & -0.418 & -0.346 &  -0.390 &  -0.438 & -0.490 & -0.543 \\
$\tilde{e}^{\rm fn}_0 $ & -0.419   & -0.349 &  -0.393 &  -0.443 &  -0.494 & -0.550 \\
Ref.~\cite{farn} ($m\to \infty$)  & -0.4252 & - &  - &  - & - & -0.5505 \\
\tableline
\tableline
\end{tabular}
\caption{Ground-state energy per spin for different values of the
$J^\prime/J$ ratio obtained with the spin-wave wave function ($e_v$),
the FN technique  ($e^{\rm fn}_0$) and the estimated 
ground-state energy ($\tilde{e}^{\rm fn}_0$) 
obtained by assuming  that the FN  error in the energy 
is weakly  size dependent  (see text).
The data for $J^\prime/J=0$ are normalized to
give the energy per spin on the kagome lattice.
Uncertainties are of the order of one unit on the last digit.
The coupled cluster method results, from Tab. I and Fig.~2 of Ref.~\cite{farn},
are also shown for comparison. }
\label{ergtab}
\end{table}

\begin{figure}
\vspace{-5mm}
\includegraphics[width=80mm]{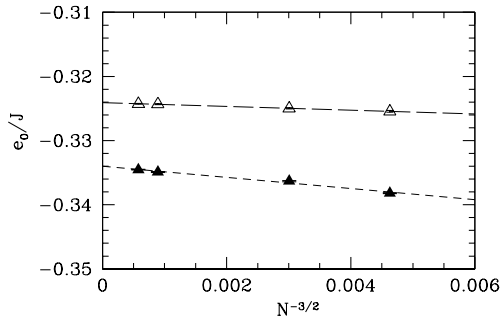}
\vspace{-10mm}
\caption{Size scaling of the ground-state energy per spin for $J^\prime/J=0.2$:
Spin-wave wave function (empty triangles), FN  (full triangles)}.
\label{ergss}
\end{figure}

A similar size-scaling analysis can be carried out to estimate the
ground-state energy per spin in the thermodynamic limit as illustrated in
Fig.~\ref{ergss}. The extrapolated energies are listed in Tab.~\ref{ergtab}.
Since the FN  energy 
error is known exactly up to the  $6\times6$ cluster, we can 
estimate the ground-state energy  in the thermodynamic limit, by adding 
 the $6\times6$  correction to the infinite-size FN  estimate.
The corresponding extrapolated values of the ground-state energy shown in the 
Tab.\ref{ergtab} represent  reasonable benchmark values for this quantity
(or at least good upper bounds), as the FN  error is expected  
to increase weakly for larger  sizes even for a good variational ansatz. 
This behavior 
has been verified up to $6\times6$, and is also very reasonable to expect in general 
for an approximate variational calculation. 

In the triangular limit, $J'/J=1$, our estimated ground-state 
energy coincides with the corresponding one 
obtained with the coupled cluster method \cite{farn} by extrapolating
in the size of the clusters  ($m\to \infty$, with  the notations  of 
Ref.~\cite{farn}). 
This value is instead slightly lower than the extrapolated  results based on
small clusters ($N \leq 36$) obtained in Ref.\cite{bernu2}, giving
$e_0=-0.5445$. 
A good agreement between the coupled cluster for $m=6$ 
and the  FN method  is also seen  for values of $J'/J$ down to
$0.2$ (Tab.~\ref{ergtab}). Furthermore,   our extrapolated energies, based 
on the error for the $6 \time 6$ cluster and the FN energy, 
remain lower  than  the $m=6$ coupled 
cluster result by a similar amount,  suggesting  that our variational approach
remains enough accurate also in this region (unfortunately the extrapolations  
$m\to \infty$ are not given in Ref.\cite{farn} for $0<J^{\prime}/J<1$).
 Instead  in the kagome limit
our variational ansatz should be 
 clearly less accurate, as the ground state is believed to be  
a spin liquid with no magnetic order in the thermodynamic limit, 
i.e. qualitatively different from   our initial variational guess given 
by Eq.~(\ref{wf}).
This may  explain why in this case  our energy estimate
is  slightly  higher than the $m\to \infty$ coupled cluster result.
We note, however, that our values reported in Tab.~(\ref{ergtab}) 
represent in general reasonable upper bounds for the energy, 
as they are obtained by a rigorous variational method such as the FN one. 

\section{Low-energy excitations}

An effective method to investigate the possibility of magnetic 
order in spin systems is to analyze the structure of the spectrum
of finite-size samples, following the strategy of Refs. 
\cite{bernu1,bernu2,lech}. The tendency toward antiferromagnetic order
in the thermodynamic limit manifests itself in finite-size clusters 
through the fact that the spin excitations with the
 lowest energies  can be described by the
effective Hamiltonian of a ``quantum top''.
Within such a description, the energy of the lowest levels in 
the different subspaces labeled by total spin $S$, can be approximated by,
\begin{equation}
E_N(S)- E_N(0) = \frac{S(S+1)}{2 I_N},
\label{pt}
\end{equation}
where $I_N$ is the inertia of the top, which is an extensive quantity.
Hence, the plots of the lowest energy levels as functions of $S(S+1)$, 
have the appearance of a ``Pisa tower'' with a slope that decreases
as $N$ increases. 
An important property of the states of the ``Pisa tower'' 
(indicated in  Ref.~\cite{bernu2} {\em quasi degenerate joint states}) is that
they belong to irreducible representations of the point group that
are compatible with the symmetry of the ordered state. In the case
of the $\sqrt{3} \times \sqrt{3}$ order, these are the 
representations labeled as $\Gamma_1, \Gamma_2, \Gamma_3$ 
of the $C_{3v}$ group, which correspond to 
$[{\bf k}=0, {\cal R}_{\pi} \Psi = \Psi, {\cal R}_{2\pi/3} \Psi = \Psi,\sigma_x \Psi = \Psi]; 
[{\bf k}=0, {\cal R}_{\pi} \Psi = -\Psi, {\cal R}_{2\pi/3} \Psi = \Psi,\sigma_x \Psi = \Psi];
[{\bf k}={\bf Q}, {\cal R}_{\pi} \Psi = \Psi, {\cal R}_{2\pi/3} \Psi = \Psi,\sigma_x \Psi = \Psi]$,
respectively. We have followed the same notation of 
Refs. \cite{bernu1,bernu2,lech}, ${\cal R}_{\phi}$ 
denoting a rotation of $\phi$,
$\sigma_x$, a reflection with respect to a mirror plane with the normal
pointing along $x$, and being ${\bf Q}=(2\pi/3,-2\pi/3)$, the corner of the
Brillouin zone of the kagome lattice.

\begin{figure}
\includegraphics[width=80mm]{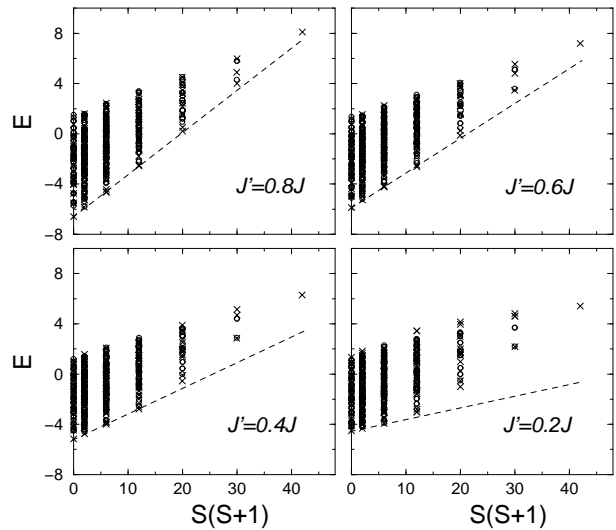}
\caption{ Energy spectra as functions of $S(S+1)$ for a cluster with $N=12$
spins, periodic boundary conditions, and $J'/J=0.8,0.6,0.4,0.2$
(left to right and top to bottom). The crosses (circles)
indicate energy levels belonging to symmetry representations
compatible (incompatible) with the $\sqrt{3}\times\sqrt{3}$ magnetic
order.
}
\label{fig7}
\end{figure}

\begin{figure}
\includegraphics[width=80mm]{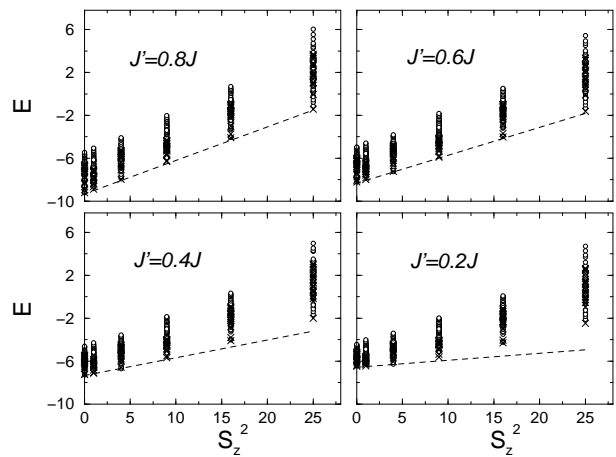}
\caption{ Energy spectra as functions of $S_z^2$ for a cluster with $N=16$
spins and twisted boundary conditions. Details are the same as
in Fig. \ref{fig7}
}
\label{fig8}
\end{figure}

\begin{figure}
\includegraphics[width=80mm]{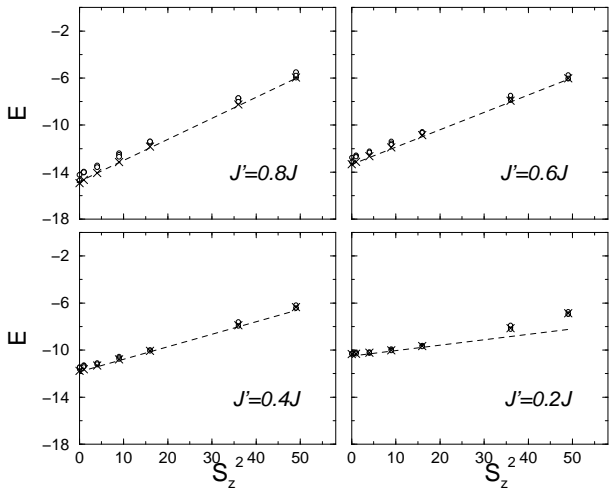}
\caption{ Lowest energy levels within the different $S_z^2$ subspaces
for $N=28$ and twisted boundary conditions.
Details are the same as in Fig.~\protect\ref{fig7}
}
\label{fig9}
\end{figure}

There are not so many finite periodic clusters, 
which can be exactly diagonalized and 
that interpolate between the
triangular and kagome lattices without frustrating the
antiferromagnetic order.
The smallest ones are $N=12$ and $N=36$, which evolve, respectively,
 toward $N=9$ and $N=27$ at the pure kagome limit ($J'=0$). 
The unit cell of the interpolating lattice
contains $2 \times 2 $ unit cells of the pure triangular lattice,
implying a reduction of the translation operations of the periodic
cluster of a factor $4$ with respect to the number of translations
at the pure triangular limit. This implies a much bigger
Hilbert space to be dealt with in the numerical procedure.
In concreteness, even exploiting all the available symmetries,
the number of states is $42035724$ for $N=36$ spins. In this cluster, 
a complete study of several excited
states within different spin sectors for several values of $J'$
is prohibitive  from the computational point of view, while the
$N=12$ cluster maybe too small. We have, therefore,
 included in the
analysis the clusters with $N=16$ and $N=28$ sites
(which evolve toward kagome clusters with  $N=12$ and $N=21$, respectively),
by introducing
twisted boundary conditions as explained in Refs. \cite{bernu2,lech}.
The latter are equivalent to rotate the local frame in 
$(\pm 2\pi/3,\mp 2\pi/3)$ at each translation along a unitary lattice vector.
This procedure restores the otherwise frustrated antiferromagnetic
order in the $x,y$ spin plane  but obviously 
breaks the spin-rotational symmetry as well as some symmetry operations of the 
point group.
For this reason, in this case,  
a law like that expressed in (\ref{pt}), with the replacement
$S(S+1)\rightarrow S^2_z$, should be
obeyed by  the quasi degenerate joint states, as the total  $S^z$ 
along the z axis,   cannot couple with the remaining  
total spin components in the clusters  with  twisted boundary conditions.
 Results are shown in Figs.~\ref{fig7},~\ref{fig8}, and \ref{fig9}
 for $N=12,16$, and $28$, respectively. Figures \ref{fig7} and \ref{fig8} 
show the full spectra, obtained
by diagonalizing  all the blocks of the Hamiltonian matrix.
In the case of $N=28$ (Fig. \ref{fig9}), we have not diagonalized the 
full Hamiltonian
but obtained the lowest eigenvalues within all the subspaces using
the Lanczos algorithm.   

\begin{figure}
\includegraphics[width=80mm]{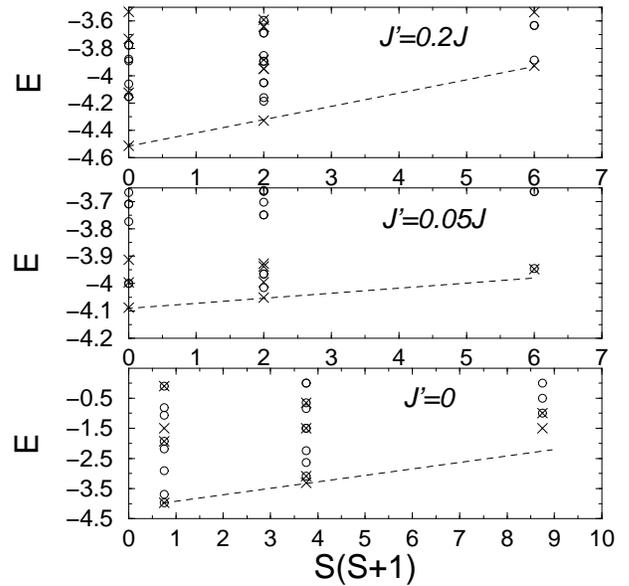}
\caption{Detail of the low-lying energy levels  close to the
kagome limit for a cluster with $N=12$ spins and periodic
boundary conditions. The lower panel
shows the behavior of the low energy levels at the kagome limit
in a cluster with $N=9$ sites. 
Details are the same as in Fig. \ref{fig7}.
}
\label{det12}
\end{figure}


%
%
%

 We 
have  plotted the states with symmetries
compatible (non-compatible) with the 
antiferromagnetic $\sqrt{3}\times\sqrt{3}$ order with crosses (circles).
The first striking feature is the fact that only a subset of the states
forming the ``Pisa tower'' in the triangular limit remains aligned
as $J'/J$ decrease. Those states build up, so to say, a ``small Pisa tower'',
whose slope decreases with $J'$.
\begin{figure}
\includegraphics[width=80mm]{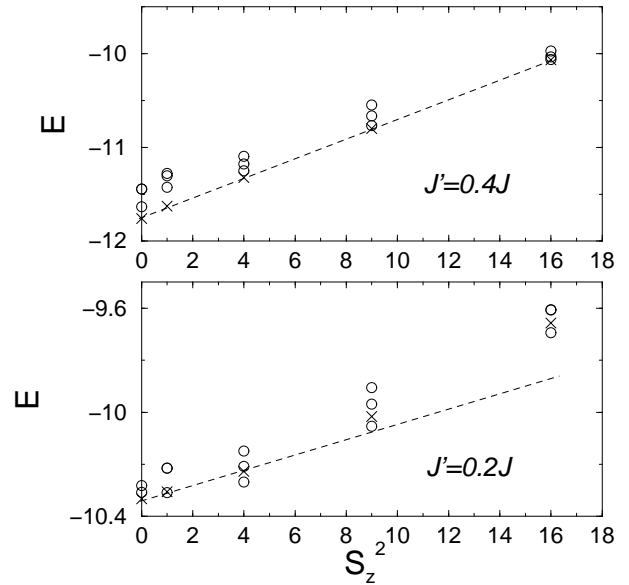}
\caption{ Detail of the low-lying energy levels  close to the
kagome limit for a cluster with $N=28$ spins and twisted
boundary conditions. 
Details are the same as in Fig. \ref{fig7}.}
\label{det28}
\end{figure}

\begin{figure}
\includegraphics[width=80mm]{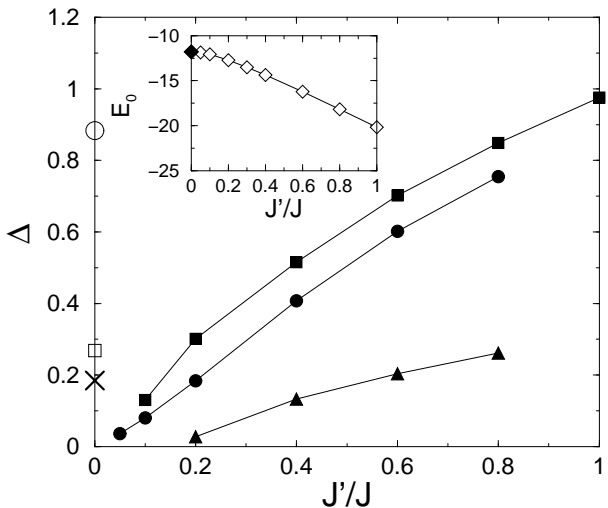}
\caption{Gap to the lowest excitation with $\Delta S_z=1$ for
$N=12$ (circles), $N=28$ (triangles), and (in the $\Gamma_1$ representation)
$N=36$ (squares).
The open circle and square at $J'=0$ indicate the
value of the spin gap for the $N=9$ and $N=27$ kagome cluster
(corresponding to the $N=12$ and $N=36$ depleted triangular clusters). 
The cross at $J'=0$
corresponds to a kagome cluster with $N=36$. 
Inset: The ground-state energy as function of $J'$ (open diamonds). The filled 
diamond indicates the ground-state energy of the 
kagome cluster with $N=27$ spins.}
\label{gappo}
\end{figure}

In order to analyze in more detail the behavior of the states
along the ``small Pisa tower'' we show zooms of the low-energy
and low-spin sector of the spectra for the $N=12$, and $N=28$
clusters in Figs. \ref{det12} and \ref{det28}, respectively. In the case of 
$N=12$, shown in Fig. \ref{det12},
this small set contains levels that belong to representations compatible
with the $\sqrt{3}\times \sqrt{3}$ order and they remain
aligned  within all the range $0.1 \leq J'/J \leq 1$.
Signatures of departure from that behavior are observed for
very low $J'/J=0.05$, where the lowest level with $S=2$
deviates from the line of the ``small Pisa tower'' while
a state with a symmetry not compatible with the antiferromagnetic
order becomes quasi degenerate with it.
In Fig. \ref{det12} the data for the related 
$N=9$ kagome lattice are also shown. In that case, a similar deviation
of the state with $S=5/2$
from the line connecting the lowest levels in the
$S=1/2$ and $S=3/2$ sectors is observed but within a scale
which is an order of magnitude larger in the energy axis.
Another contrasting feature that comes from the comparison 
between $J'=0$ and  $J'\neq 0$ is that
in the pure kagome limit, there are two states with $S=1/2$
(one of them belonging to the manifold of the degenerate ground state)
 whose symmetries are not compatible with the 
$\sqrt{3}\times \sqrt{3}$ order that have energies within the spin gap.
Instead for $J^\prime/J=0.05$ the spin gap is clean from such states. 
In addition, the gap to the lowest excitation with
$\Delta S=1$ does not evolve with continuity for $J^\prime/J\to 0$ 
to the corresponding gap in the kagome limit (see also Fig.~\ref{gappo}).

Similar remarks apply to the 16- and 28- sites clusters
shown in Figs.~\ref{fig8} and \ref{fig9}. In these cases, see e.g., 
Fig.~\ref{det28},
the ``small Pisa Tower'' contains a larger number of states 
but the deviation 
from a perfect alignment is observed already at $J'/J \leq 0.2$,
where states within subspaces not compatible with the 
$\sqrt{3}\times\sqrt{3}$ order appear at low energies. In particular,
for $S_z=1$ one of those states is almost degenerate with the one belonging
to the Pisa tower while for $S_z=2$ it is well below it.
As in the $N=12$ cluster, also for $N=16$ and 28 
the gap to the first spin excitation monotonically
decreases with $J^\prime$ (Fig.~\ref{gappo}).

To complete the analysis, we have 
computed the ground-state energy  ($E_0$) and
the lowest eigenvalue in the subspace with $S=1$ within the representation $\Gamma_1$ 
of $C_{3v}$ ($E_{S=1,\Gamma_1}$), in the periodic lattice
with $N=36$ sites. The latter does not actually
 correspond to the lowest energy $S=1$ excitation in the pure triangular 
limit, which is in the subspace corresponding to $\Gamma_2$.
In any case, the energy difference 
$\Delta=E_{S=1,\Gamma_1} - E_0$
is, in general, an upper bound for the spin gap of the cluster in all the 
range $0<J^\prime/J\leq 1$.
The behavior of $\Delta$ as a function of $J'$ is
 shown in Fig.~\ref{gappo}. 
As in the clusters previously considered, the gap decreases
rapidly as $J'$ decreases, and, in particular, 
it does not evolve continuously for $J'\to 0$  towards the value
of the corresponding kagome lattice  
($N=27$, indicated with an open square in Fig.~\ref{gappo}).
Instead, it tends to a value which is even smaller than the
magnitude of the spin gap for the $N=36$ kagome lattice
(indicated by a cross in the same figure). This is in 
contrast with the behavior exhibited by the ground-state energy, which
evolves smoothly toward the value of the $N=27$ kagome
lattice (see the inset of Fig.~\ref{gappo}).

The behavior of the spin gap and the spectra strongly suggests the
closing of the spin-gap in the thermodynamic limit within 
the whole range of $0<J'/J\leq1$. In fact, such gap is known to close 
in the thermodynamic limit at the triangular point ($J'/J=1$)
\cite{bernu1,bernu2,luc} and  it decreases systematically by decreasing $J'$
in all the cluster considered
as the low-energy scale for the lowest-spin excitations is set by $J^\prime$,
i.e., by the states of the ``small Pisa tower''.
This can be understood in a simple uncorrelated framework where for
$J^\prime< J$ the lowest spin-excitations are clearly
obtained through a spin-flip on a non-kagome site, with an energy
cost $~6J^\prime$, compared to a cost $~4J+2J^\prime$ 
for a spin flip on a kagome site.
As the number of non-kagome spins is just a fraction of 
the total number of sites, on a finite cluster 
such a mechanism would apply only to the lowest spin excitations:
hence the reduction of the number of states belonging to 
the ``Pisa tower'' with respect to the triangular case.

This simple picture immediately suggests that the nature 
of the spin excitations is intrinsically different for finite
$J^\prime$ and  in the pure kagome limit. In the latter case, in fact,
the spins on the non-kagome sites do not belong
to the Hilbert space of the model and such low-energy spin excitations are not
possible. For this reason by turning on the $J^\prime$, the spin gap
has a finite discontinuity and it is reasonable to expect the model to
be unstable under the perturbation introduced by $J^\prime$ bonds.
This is also confirmed by the analysis of the low-energy 
excitations for small values of $J^\prime/J$ (see Fig.~\ref{det12})

One step further in the analysis of the nature of the ground state
leads to the question whether or not the latter remains
ordered down to the kagome limit. 
The structure of the spectra of the small $N=12$ cluster 
indicates that could well be the case, while those of $N=16$ and
$N=28$ suggests some change in the nature of the ground state
around $J'/J \sim 0.2 $. It is worth recalling, however, 
that the results for the
two latter clusters have been obtained by using twisted boundary conditions
which could lead to weaker signatures of ordering.
On  the other hand, in two dimensions, 
it seems difficult to conceal the possibility of
the closing of the spin gap with the absence of some kind of magnetic order.
A simple geometrical analysis reveals that the non-kagome spins,
which we have already identified as the responsible of the lowest-energy
excitations, form themselves a triangular lattice with a cell parameter
twice the size of the one of the usual triangular lattice.
Therefore, a possible scenario for the evolution of the ground state as
$J'$ decreases could be that at some point a crossover takes place from a
magnetic order in the usual $\sqrt{3}\times\sqrt{3}$ pattern to a
magnetic order with a similar pattern but in the triangular lattice
of the non-kagome spins. Both kinds of order are commensurate and the
ground state could undergo a smooth evolution from one to the other.

\section{Summary and Conclusions}
We have investigated the low energy properties
of a $J'-J$ triangular lattice that interpolates
between the usual triangular ($J'/J=1$) and kagome ($J'=0$)
lattices. 
To this end, we have used
a variational approach based on a FN wave function accurately
describing the ground state in the triangular ($J^\prime/J=1$) 
limit, and exact diagonalization techniques.

We have analyzed the quality of the approximation to the exact
ground state, provided
by the FN technique in a periodic cluster with $N=36$ sites and 
then extended the calculation up to $N=144$ sites by using quantum Monte Carlo. 
We have found that such a wave function, describing a state with
a $\sqrt{3}\times\sqrt{3}$ N\`eel ordered phase, is 
very close to the exact one for  $0.2 \lesssim J^\prime/J \leq 1$ and
that in this range its accuracy is almost independent 
of the $J^\prime/J$ ratio. Consistently, the ground-state
expectation value of the antiferromagnetic order parameter 
remains approximately constant down to $J^\prime/J\simeq 0.2$. 
Below this value
the order parameter is suppressed and the quality of the 
FN wave function degrades, suggesting a change in the nature of
the ground-state.
This is also confirmed by the analysis of the 
low-energy spectra on small clusters, showing some signatures of
instability of N\`eel ordering for $J^\prime/J\lesssim 0.2$.
However, at the same time, such analysis also indicates quite clearly
that the low-energy scale for spin excitations is set by $J^\prime$
and that in particular the spin-gap has a finite discontinuity at 
$J^\prime/J=0$, the triplet excitations  being gapless 
for any non zero value of 
$J^\prime/J$.
This would imply that the kagome disordered phase is
unstable against a slight perturbation tending to
restore the $z=6$ coordination number of the triangular lattice.
A possible scenario is that within
the large amount of singlets that are quasi degenerate with
the ground state in the kagome clusters, 
the one corresponding to the  $\sqrt{3}\times\sqrt{3}$
state is favored by some kick produced by $J'$.

Our results are  in disagreement with the predictions of the spin-wave
theory of Ref.~\cite{kago1}, indicating a progressive reduction
of the antiferromagnetic ordering for $J^\prime/J<1$ and a complete melting
of the $\sqrt{3}\times\sqrt{3}$ order for $(J^\prime/J)_c\simeq0.2$. 
Instead, our conclusions are closer to those of the recent
coupled cluster treatment of Farnell {\em et al.} \cite{farn} providing
evidence for instability of the antiferromagnetic 
order very close or possibly at 
the kagome point ($(J^\prime/J)_c=0.0\pm 0.1$), and the existence for small
$J^\prime/J$ of a regime whose correlations are very different to those
of the triangular antiferromagnet. 
Further investigations are necessary to clarify the nature of this regime.

\section{Acknowledgments}
L. A. thanks Prof. Fulde for his hospitality as well as the 
the Alexander-von-Humboldt Stiftung and CONICET, Argentina for the
support. This work was partially supported by INFM-PRA MALODI, 
and by NSF under Grant No. DMR02-11166 (L.C.).

\end{document}